\documentclass[10pt,twocolumn,superscriptaddress,english,10pt,prl,showpacs,floatfix,aps]{revtex4-1}
\usepackage[latin9]{inputenc}
\usepackage{amsthm}
\usepackage{amsmath}
\usepackage{amssymb}
\usepackage{esint}
\usepackage{graphicx}
\usepackage{upgreek}
\usepackage{xspace}

\makeatletter

\@ifundefined{textcolor}{}
{%
 \definecolor{BLACK}{gray}{0}
 \definecolor{WHITE}{gray}{1}
 \definecolor{RED}{rgb}{1,0,0}
 \definecolor{GREEN}{rgb}{0,1,0}
 \definecolor{BLUE}{rgb}{0,0,1}
 \definecolor{CYAN}{cmyk}{1,0,0,0}
 \definecolor{MAGENTA}{cmyk}{0,1,0,0}
 \definecolor{YELLOW}{cmyk}{0,0,1,0}
}

\usepackage{soul}
\usepackage{braket}
\usepackage[backref=false,
bookmarksnumbered=true,
bookmarks=true,
bookmarksopen=true,
colorlinks=true,
citecolor=blue,
linkcolor=blue,
anchorcolor=green,
urlcolor=blue,unicode=false]{hyperref}

\let\baraccent=\= 
\renewcommand{\=}[1]{\stackrel{#1}{=}} 

\DeclareMathOperator{\Tr}{Tr}

\newcommand{\didv}{d$I/$d$V$\xspace}
\newcommand{\etal}{\textit{et al.}\xspace}

\newcommand{\DeltaT}{\ensuremath{\Delta_\mathrm{tip}}\xspace}
\newcommand{\DeltaS}{\ensuremath{\Delta_\mathrm{sample}}\xspace}
\newcommand{\Fig}[1]{Fig.~\ref{fig:#1}}
\newcommand{\Figure}[1]{Figure~\ref{fig:#1}}
\newcommand{\MainFig}[1]{Fig.~#1}

\DeclareMathOperator{\unitspace}{\xspace}
\DeclareMathOperator{\unm}{\unitspace\mathrm{nm}}
\DeclareMathOperator{\upm}{\unitspace\mathrm{pm}}
\DeclareMathOperator{\uV}{\unitspace\mathrm{V}}
\DeclareMathOperator{\umV}{\unitspace\mathrm{mV}}
\DeclareMathOperator{\umuV}{\unitspace\mathrm{\mu V}}

\DeclareMathOperator{\umeV}{\unitspace\mathrm{meV}}
\DeclareMathOperator{\umueV}{\unitspace\mathrm{\mu eV}}

\DeclareMathOperator{\upA}{\unitspace\mathrm{pA}}
\DeclareMathOperator{\unS}{\unitspace\mathrm{nS}}

\DeclareMathOperator{\uK}{\unitspace\mathrm{K}}

\DeclareMathOperator{\uHz}{\unitspace\mathrm{Hz}}
\DeclareMathOperator{\uAA}{\unitspace\mathrm{\AA{}}}

\makeatother

\usepackage{babel}

\begin{document}

\title{End states and subgap structure in proximity-coupled chains of magnetic adatoms}

\author{Michael Ruby}
\affiliation{\mbox{Fachbereich Physik, Freie Universit\"at Berlin, 14195 Berlin, Germany}}

\author{Falko Pientka}
\affiliation{\mbox{Dahlem Center for Complex Quantum Systems and Fachbereich Physik, Freie Universit\"at Berlin, 14195 Berlin, Germany}}

\author{Yang Peng}
\affiliation{\mbox{Dahlem Center for Complex Quantum Systems and Fachbereich Physik, Freie Universit\"at Berlin, 14195 Berlin, Germany}}

\author{Felix von Oppen}
\affiliation{\mbox{Dahlem Center for Complex Quantum Systems and Fachbereich Physik, Freie Universit\"at Berlin, 14195 Berlin, Germany}}

\author{Benjamin W.\ Heinrich}
\affiliation{\mbox{Fachbereich Physik, Freie Universit\"at Berlin, 14195 Berlin, Germany}}

\author{Katharina J.\ Franke}
\affiliation{\mbox{Fachbereich Physik, Freie Universit\"at Berlin, 14195 Berlin, Germany}}

\begin{abstract}
A recent experiment [Nadj-Perge \etal, Science {\bf 346}, 602 (2014)] provides evidence for Majorana zero modes in iron (Fe) chains on the superconducting Pb(110) surface. Here, we study this system by scanning tunneling microscopy using superconducting tips. This high-resolution technique resolves a rich subgap structure, including zero-energy excitations in some chains. We compare the symmetry properties of the data under voltage reversal against theoretical expectations and provide evidence that the putative Majorana signature overlaps with a previously unresolved low-energy resonance. Interpreting the data within a Majorana framework suggests that the topological gap is significantly smaller than previously believed. Aided by model calculations, we also analyze higher-energy features of the subgap spectrum and their relation to high-bias peaks which we associate with the Fe d-bands. 

\end{abstract}


\maketitle

Building on advances in nanofabrication \cite{francesci}, engineering topological phases by proximity in superconducting hybrid structures has come within reach of current experiments. A major motivation for realizing such phases are their non-abelian Majorana quasiparticles \cite{review1,review2,review3}, and their subsequent applications. The underlying topological superconducting phases can be realized in one-dimensional (1d) helical liquids contacted by conventional $s$-wave superconductors \cite{fu09,lutchyn10,oreg10,nadj-perge13,li2014}. Among the most promising platforms studied in experiment are semiconductor nanowires \cite{mourik12,das12,lund,rokhinson12,harlingen,churchill13}, edges of two-dimensional topological insulators \cite{hart,pribiag}, and chains of magnetic adatoms \cite{yazdani,pawlak2015afm}. While the proximity coupling to a superconductor is needed to induce a gap protecting the topological phase, it also has more subtle consequences. Magnetic interactions mediated by the superconductor can stabilize magnetic order in the 1d system \cite{braunecker13,vazifeh13,klinovaja13,kim14}. 
Conversely, the spin structure may affect the superconductor. This is particularly apparent for adatom chains, where a band of subgap Shiba states \cite{yu,shiba,rusinov,balatsky} may strongly modify the low-energy properties of the system \cite{nadj-perge13,nakosai,pientka13,ojanen,kotetes,sau} and
possibly induce trivial zero-energy features at the chain end \cite{sau15}. At strong coupling, the 1d states bleed substantially into the superconductor, reducing the effective coherence length at low energies \cite{peng2015}. 

Nadj-Perge {\em et al}.\ \cite{yazdani} recently provided intriguing evidence for Majorana states in Fe chains on Pb(110). Here, we present data on the same system employing scanning tunneling microscopy/spectroscopy (STM/STS) with superconducting tips (see also \cite{yazdani}). We show that the use of superconducting tips not only provides enhanced resolution of the subgap structure, but also allows for additional consistency checks on the interpretation of the data in terms of Majorana quasiparticles. Our observations indicate a subgap spectrum comprising a flat Shiba band and strongly dispersing Fe states. An interpretation in terms of Majorana states suggests that the induced gap is considerably smaller than previously believed. 

We carried out the experiments in a \textsc{Specs} JT-STM at a temperature of $1.1\uK$. Cycles of sputtering and annealing of a Pb(110) single crystal ($T_\textrm{c}=7.2\uK$) resulted in an atomically flat and clean surface. We employed Pb-covered superconducting tips (see Ref.\ \cite{franke11} for the preparation procedure), which provide a resolution beyond the Fermi-Dirac limit~\cite{ji08,bheinrich13,ruby14} (in our measurements: $\simeq70\umueV$). Fe chains were prepared by e-beam evaporation from an iron rod (99.99\% purity) onto the clean surface at room temperature, similar to Ref.~\cite{yazdani}. Without further annealing, we obtained chain lengths of up to $\simeq10\unm$ (measured between chain end and intervening Fe cluster). Single adatoms and dimers were prepared by e-beam evaporation onto the cold sample in the STM ($T<10\uK$) with a density of $\approx350$ adatoms per $100\times100\unm^2$. The differential conductance \didv as a function of sample bias was recorded using standard lock-in technique at $912\uHz$ (subgap spectra: bias modulation $V_\textrm{mod}=15\umuV_\mathrm{rms}$, setpoint $V= 5\umV$, $I=250\upA$; large-bias spectra: $V_\textrm{mod}=2\umV_\mathrm{rms}$, $V=2\uV$, $I=850\upA$).

The \didv-signal measures a convolution of the densities of states (DOS) of tip and sample when the tunneling rate is slower than the quasiparticle relaxation of the subgap states~\cite{ruby15}. This is the case for all measurements presented in this paper, as the differential conductance varies linearly with junction transparency. STS with a superconducting tip shifts all sample peaks by $\pm\DeltaT$. The coherence peaks then appear as resonances at $eV=\pm(\DeltaT+\DeltaS)$. Subgap states of the substrate of energy $\varepsilon$ appear as \didv resonances at voltages $\pm\left(\DeltaT+\varepsilon\right)$ due to tunneling between the occupied (unoccupied) DOS singularity of the tip and the unoccupied (occupied) subgap state. At biases below $\DeltaT$, resonances occur at $\pm\left(\DeltaT-\varepsilon\right)$ which result from thermally excited quasiparticles in substrate or tip. At our experimental temperature, this effect is limited to small energies $\varepsilon$. Accordingly, zero-energy Majorana states are signaled by resonances at $eV=\pm \DeltaT$ \cite{yazdani}, and we thus pay particular attention to biases near this threshold.

\begin{figure}[tb]
\includegraphics[width=0.48\textwidth]{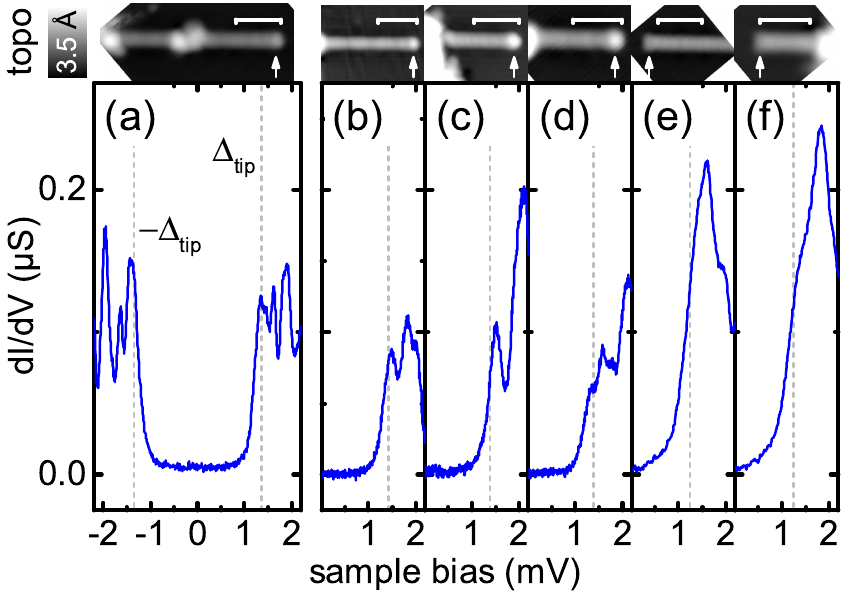}
\caption{
\didv-spectra recorded at the end of six different chains (small arrows mark position on the chain). Chains in (a-d) are terminated by a small cluster, in (e,f) they have a sharp cut-off. The energy resolution in (e,f) is reduced (width of BCS resonance: $\simeq330\umuV$), due to non bulk-like superconductivity of the tip. \DeltaT in $\umV$: (a) $1.36$, (b) $1.42$, (c,d) $1.38$, (e,f) $1.24$. Chain lengths measured between chain end and cluster onset in $\unm$: (a) 13.9, (b) 9.5, (c) 6.2, (d) 6.0, (e) 7.7, (f) 4.0.
Scale bars correspond to $4\unm$. For spectra of (b-f) at negative bias see Supplementary Material~\cite{supplementary}.
}
\label{fig:1}
\end{figure}

\Figure{1} shows subgap \didv-spectra, recorded at the termination of six independent Fe chains. The chains in \Fig{1}(a,b,d) exhibit a clear zero-energy signature at a bias of $+\DeltaT$, which has been interpreted as a fingerprint of a Majorana bound state \cite{yazdani} (for the determination of the value of \DeltaT, see Supplementary Material~\cite{supplementary}). The zero-energy feature is accompanied by two resonances at higher energy. In contrast, the chain in \Fig{1}(c) does not exhibit a clear peak at $+\DeltaT$ and also differs in the remaining subgap structure, with a resonance at around $1.52\umV$. \Figure{1}(e) and (f) show data for chains without a protrusion at their ends. These appear rarely and we can only provide data with reduced resolution ($\simeq330\umuV$) due to inferior tip preparation. These chains also lack an unambiguous signature of a zero-energy resonance, perhaps masked by the larger resonance which contributes considerable spectral intensity at \DeltaT. 

\Figure{1}(a) shows that the peaks at opposite biases $\pm \DeltaT$ differ substantially in intensity. The same is observed in all other chains \cite{supplementary}. This is in contrast to expectations for Majorana peaks, which should be symmetric, as a Majorana state has identical electron and hole wavefunctions~\cite{yang2}. This indicates that these peaks originate at least partially from trivial subgap states near zero energy. For a more detailed analysis, we focus on the chains shown in \Fig{1}(a,b).

\begin{figure}[bt]
\includegraphics[width=0.48\textwidth]{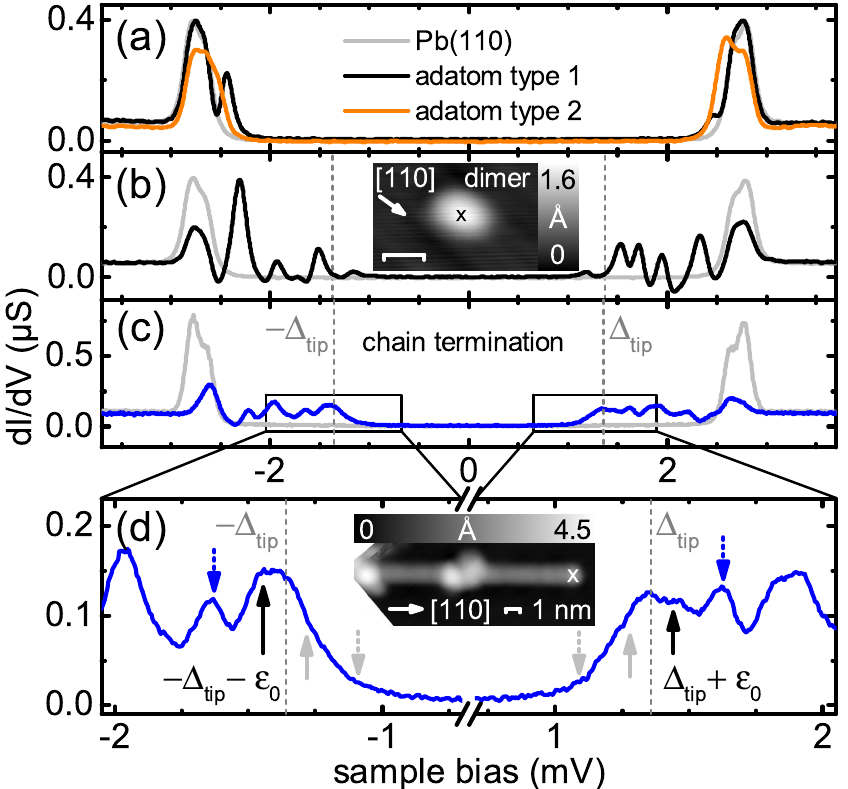}
\caption{
(a) \didv-spectra of single Fe adatoms (two types: 1 and 2). Both show a single, loosely bound Shiba-resonance close to the gap edge. They differ in apparent height by $\simeq20\upm$ at $50\umV$, $50\upA$.
(b) Dimers show a variety of Shiba states with the lowest energy resonance at $150\umueV$. Scale bar corresponds to $1\unm$. $\DeltaT=1.39\umV$.
(c) \didv-spectrum at the end of the chain (blue), compared to bare Pb(110) (gray). The chain is the same as shown in \Fig{1}(a).
(d) Zoom on the spectrum in (c). The subgap spectrum shows a manifold of Shiba resonances. A peak at $+\DeltaT$ may be the fingerprint of a Majorana bound state. Low energy states lie at $\varepsilon_0\simeq80\umuV$ (black solid arrow) and $\varepsilon_1\simeq270\umuV$ (blue dashed arrow). The energies of the expected thermal resonances of low intensity are marked by their gray equivalents below $\DeltaT=1.36\umV$.
}
\label{fig:2}
\end{figure}

It is interesting to contrast the spectra for chains with those for individual Fe adatoms and dimers. \Figure{2}(a) shows the \didv-signal of two species of single adatoms with different apparent heights. For both types, we observe a single shallow Shiba state with energies $1.1\umeV$ (type 1) and $1.2\umeV$ (type 2), respectively. In contrast, the dimer shown in \Fig{2}(b) exhibits a richer subgap structure with a series of resonances reaching as low as $\simeq150\umueV$. We find that the subgap spectrum varies in detail between different dimers, depending on interatomic distance, angle, and adsorption site~\cite{supplementary}. This demonstrates strong coupling of Shiba states which can ultimately lead to the formation of Shiba bands in adatom chains.

\Figure{2}(c) provides the data of \Fig{1}(a) over a wider voltage range, with a zoom-in on the voltage range near \DeltaT shown in \Fig{2}(d). In addition to the peak at $+\DeltaT$ and a faint shoulder at $-\DeltaT$, there is a nearby subgap peak at $\varepsilon_0\simeq80\umuV$ (black solid arrows). These combine into a plateau-like structure near and just above $\pm\DeltaT$. The data are also consistent with corresponding thermal resonances at $\pm\left(\DeltaT-\varepsilon_0\right)$ (gray arrows). The superposition with a low-energy subgap resonance may explain the asymmetric peaks in \Fig{1}(a). While a Majorana peak must be symmetric, conventional subgap resonances can be asymmetric, reflecting the asymmetry between its electron and hole wavefunctions. 

Further subgap peaks occur at higher energies, the next higher one at $\varepsilon_1\simeq270\umuV$ (blue dashed arrows in \Fig{2}(d)). Similar peaks were identified in Ref.~\cite{yazdani} as the coherence peaks of the induced topological gap (estimated at $200-300\umueV$). In view of the lower-energy peaks, this interpretation seems implausible for our chains. Instead, a Majorana-based interpretation would suggest that the lowest non-zero energy peak originates from the topological gap or is shifted above the topological gap by size quantization. This suggests that the topological gap is comparable or smaller than $\simeq80\umueV$. 

\begin{figure}[tb]
\includegraphics[width=0.48\textwidth]{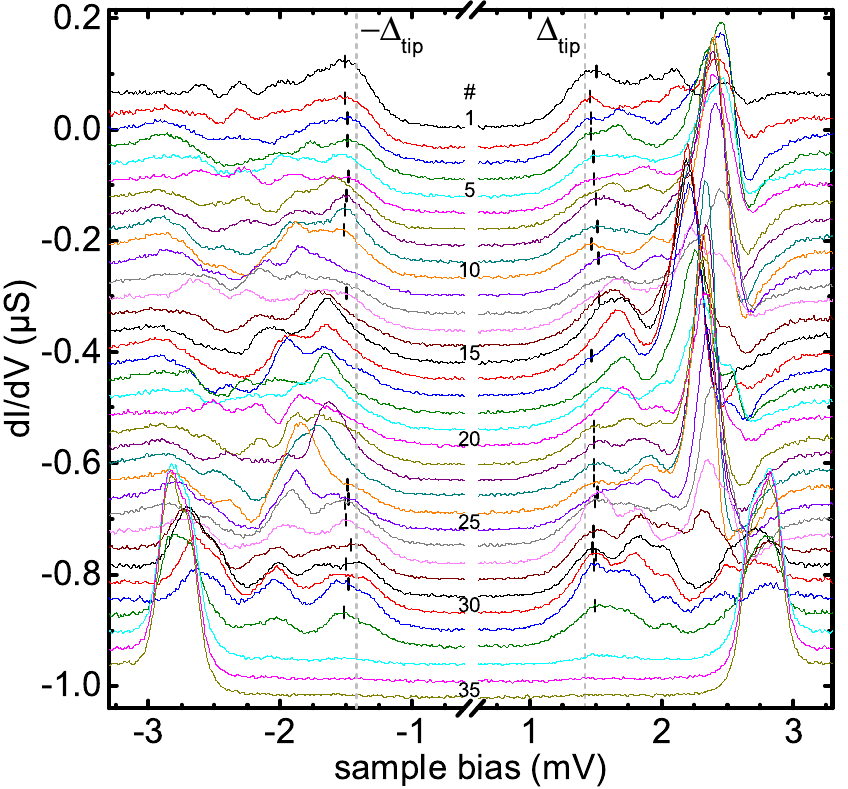}
\caption{
Spatially resolved \didv-spectra along chain (b) in \Fig{1}, going from the onset of the Fe cluster (\#1) to the bare surface (\#35).
A manifold of Shiba bands dominates the subgap structure.
At the chain end (\#29, same as \Fig{1}(b)) and the Fe cluster (\#1) a zero-energy resonance around $\pm\DeltaT$ is visible. The lowest non-zero energy resonance has $\varepsilon_0\simeq80\umueV$ in (\#29), and is modulated along the chain. As a guide to the eye low-energy resonances below a bias of $\pm\left(\DeltaT+100\umuV\right)$ are marked by small ticks. The higher energy peaks at $\simeq270\umuV$ shift in energy along the chain.
Offset for clarity: $-30\unS$/spectrum. Distance between spectra: $0.33\unm$. $\DeltaT=1.42\umV$.
}
\label{fig:3}
\end{figure}

While the Majorana states should be localized at the chain end, the topological gap is a property of the bulk spectrum and should be observable throughout the entire chain. \Figure{3} shows spatially resolved \didv-traces of chain (b) in \Fig{1}. At the end of the chain (\#29) [and the onset of the Fe cluster (\#1)], there are peaks at $\pm\DeltaT$ as well as low-energy subgap resonances with $\varepsilon_0\simeq80\umueV$, similar to the chain shown in \Fig{2}. The peaks at $\pm\DeltaT$ are again asymmetric. But while the zero-energy resonance is observed only at the end of the chain, the low-energy resonance at $\varepsilon_0$ is indeed observable elsewhere along the chain as indicated by ticks in \Fig{3}. Notice, however, that the apparent energy of the resonance and its intensity varies widely along the chain.

\begin{figure*}[tb]
\includegraphics[width=0.96\textwidth]{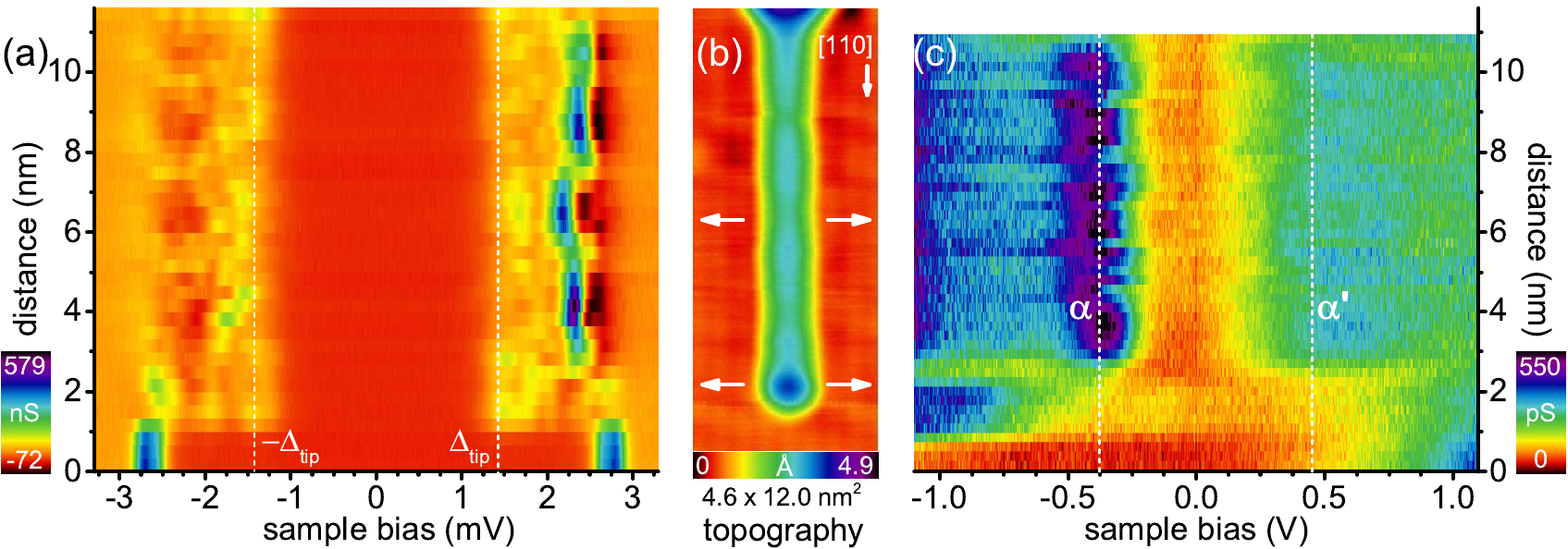}
\caption{
False-color plot of the \didv-spectra of (a) the subgap (same spectra as in \Fig{3}) and (c) the d-band structure, aligned with the topography in (b).
The subgap structure (a) exhibits variations of the Shiba peak energies and intensities, as well as modulations of the low-energy resonance at $\varepsilon_0\simeq80\umueV$ along the chain. A zero-energy resonance is found at both chain terminations.
Resonances linked to the d-band structure (c) vary around $-380\umV$ ($\alpha$), and $450\umV$ ($\alpha'$) along the chain.
}
\label{fig:4}
\end{figure*}

Many of the spectra are dominated by a strong subgap resonance at a bias of $\simeq 2.3\umV$. A false-color plot of the same \didv-spectra reveals that its intensity oscillates with a period of $\simeq 2\unm$ and shifts slightly to lower energy in the center of the chain [see \Fig{4}(a)]. The lower-energy resonances near $\DeltaT$ do not show such a clear periodicity. Interestingly, the variations appear correlated with the topography of the chain, as shown next to the false-color plot in \Fig{4}(b). Apparent height and width of the chain show variations with a similar period of $\simeq 2\unm$, in agreement with Ref.~\cite{yazdani}. 

Further insight can be gained from a false-color plot of \didv-spectra along the chain in a larger bias range between $\pm1.1\uV$ [see \Fig{4}(c)]. In the interior of the chain we observe two prominent features: a narrow resonance $\alpha$ around $-380\umV$ and a broader resonance $\alpha'$ around $450\umV$. Similar resonances are present for all chains shown in \Fig{1}, and are in agreement with Ref.~\cite{yazdani}. These resonances decrease in intensity and finally disappear at the end of the chain, or close to the Fe cluster, respectively (see additional traces in \cite{supplementary}). We interpret these resonances as van-Hove singularities of the d-bands of the Fe chain. The simultaneous disappearance of $\alpha$ and $\alpha'$ suggests that these are the upper and lower edge of the same band crossing the Fermi level. 
Interestingly, resonance $\alpha$ shifts with the above observed periodicity of about $2\unm$~\cite{supplementary}.

\begin{figure}[t]
\includegraphics[width=0.48\textwidth]{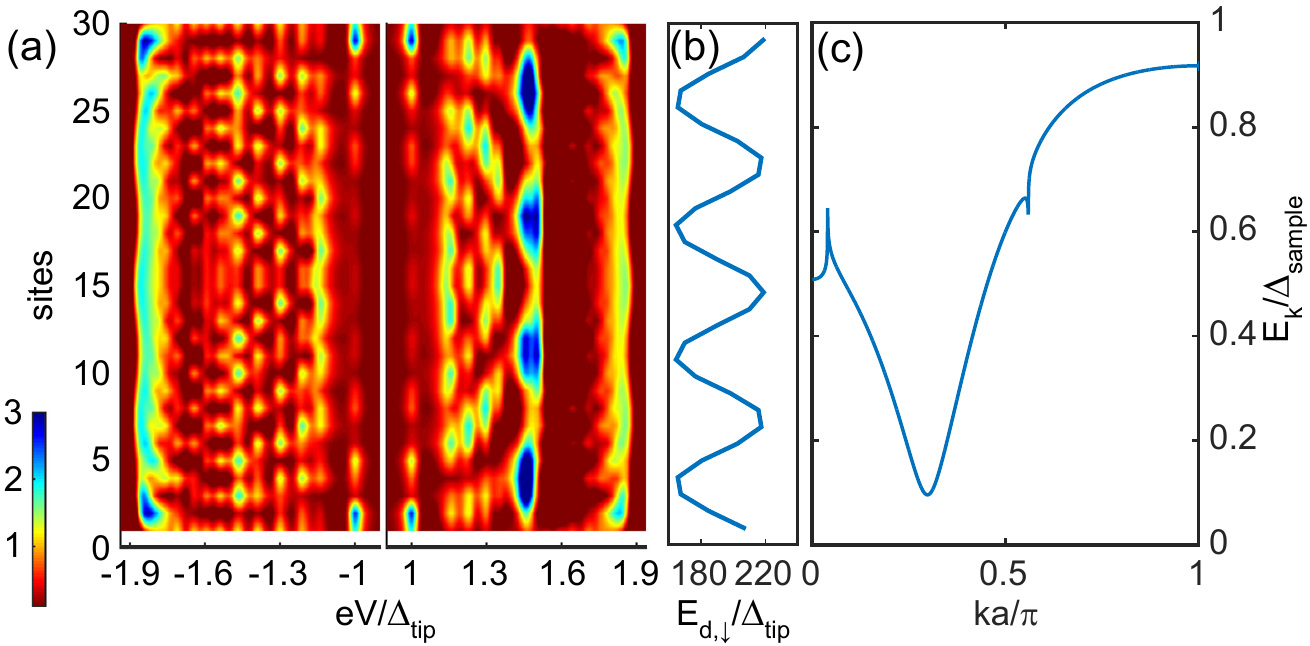}
\caption{
Numerical results for a chain of $30$ sites:
(a) Color plot of the differential conductance along the chain at subgap energies for a superconducting tip.
(b) Spatially varying on-site energies of impurity levels.
(c) Bulk subgap band structure of an impurity chain on the surface of an $s$-wave superconductor.  
We used $\DeltaS/\DeltaT=0.958$. See \cite{supplementary} for other parameters.
}
\label{fig:5}
\end{figure}

To understand the origin of the variations in the d-bands and the peaks in the subgap spectra, we have performed model calculations for a chain of Anderson impurities coupled to an $s$-wave superconductor with spin-orbit coupling, based on Ref.\ \cite{peng2015}. We account for the modulations by including a potential which varies along the chain and reflects the local environment of the adatoms. We choose parameters such that one band crosses the Fermi level with band edges corresponding to $\alpha$ and $\alpha'$ and we assume strong adatom-substrate coupling. Following \cite{peng2015}, we calculate the subgap local density of states from a mean-field treatment of the impurity chain (see \cite{supplementary}). \Figure{5}(c) shows the subgap band structure of an infinite chain. It comprises two flat parts, which can be viewed as Shiba states in the superconducting substrate around $ka\simeq 0$ and $ka\simeq \pi$, and a V-shaped dip that originates from the strongly dispersing impurity states. For direct comparison with the data, we compute the differential conductance. As temperature exceeds the typical energy separation between subgap levels, we assume efficient quasiparticle relaxation which results in dominant single-particle tunneling. In addition, we model the experimental resolution by introducing a broadening in the tip density of states. Note that these conditions preclude the observation of a quantized Majorana peak height \cite{yang2}. \Figure{5}(a) shows the differential conductance of a finite chain for subgap energies, including a spatially varying potential [shown in (b)] to model the variation of the d-level resonances in \Fig{4}(c). The numerical results are consistent with key features of the experimental data: (i) The Majorana bound state has a short decay length of a few lattice sites as a consequence of the strong chain-substrate coupling \cite{peng2015}. (ii) Prominent peaks at $eV=\pm1.5\,\DeltaT$ and $\pm1.9\,\DeltaT$ signal the van-Hove singularities of the Shiba band. Their intensity modulations are correlated with the potential landscape of the impurity atoms. Here the effect of the corrugation is most visible, as the Shiba energy explicitly depends on the energy of the impurity level. (iii) The induced gap varies along the chain on atomic scales but is uncorrelated with the potential landscape. Indeed, at strong coupling the induced gap only depends on the substrate gap and spin-orbit interaction and is insensitive to details of the impurities. The fluctuations originate from finite-size quantization which is most visible at low energies due to the low density of states in the V-shaped dip of the band structure [\Fig{5}(c)].

Motivated by Ref.~\cite{yazdani}, we investigated the subgap spectra and possible Majorana signatures of Fe chains on a superconducting Pb(110) substrate by scanning tunneling spectroscopy. Using superconducting tips, a Majorana state is expected to appear as a pair of resonances at $\pm\DeltaT$ with symmetric intensities. We associate the absence of this symmetry in the data with a nearby low-energy subgap resonance at $80\umueV$. Within a Majorana framework, it is natural to interpret this additional resonance as the coherence peak of the induced topological gap which would then be significantly smaller than previously believed. We show by model calculations that such an interpretation is in principle consistent with our observations. However, a conclusive confirmation of Majorana end states in chains of magnetic adatoms would be greatly facilitated by experiments at considerably lower temperatures. Using superconducting tips at temperatures well below the induced gap might even provide access to the elusive conductance quantization of Majorana states \cite{yang2}.

\begin{acknowledgments}
We acknowledge financial support by the Deutsche Forschungsgemeinschaft through SFB 658 and FR2726/4 (KF),
as well as SPP 1285 and SPP 1666 (FvO), also by an ERC grant NanoSpin (KF) and the Helmholtz Virtual Institute {\em New States of Matter and Their Excitations} (FvO).
\end{acknowledgments}

\clearpage

\renewcommand{\theequation}{S\arabic{equation}}

\onecolumngrid

\section*{\Large{Supplementary Material}}

\section{\texorpdfstring{{\protect \didv}}~-spectra of the subgap structure of six different chains}

\begin{figure}[h]
\includegraphics[width=\columnwidth]{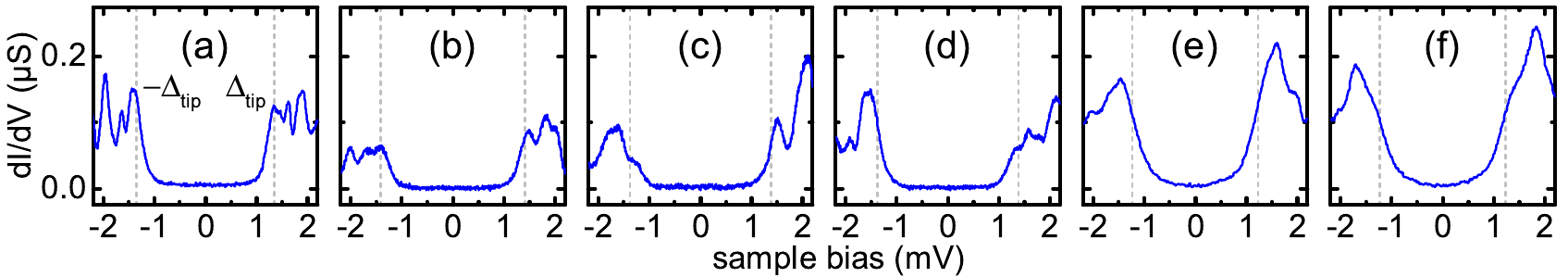}
\caption{
\didv-spectra recorded at the end of six different chains. 
Same spectra as in \MainFig{1}(a-f) of the main text, but here we present both positive and negative bias side. Chains in (a-d) are terminated by a small cluster, in (e,f) they have a sharp cut-off. The energy resolution in (e,f) is reduced (width of BCS resonance: $\simeq330\umuV$), due to non bulk-like superconductivity of the tip. \DeltaT in $\umV$: (a) $1.36$, (b) $1.42$, (c,d) $1.38$, (e,f) $1.24$. Chain lengths measured between chain end and cluster onset in $\unm$: (a) 13.9, (b) 9.5, (c) 6.2, (d) 6.0, (e) 7.7, (f) 4.0.
Setpoint: $5\umV$, $250\upA$. Lock-in: $15\umuV_\mathrm{rms}$, $912\uHz$.
}
\label{Sfig:Fig1Full}
\end{figure}

In \MainFig{1}(a-f) of the main text we show \didv-spectra at positive bias voltage of the subgap structure at the end of six different chains. For the sake of completeness, we show here the same spectra, but both positive and negative bias side [Fig.~\ref{Sfig:Fig1Full}(a-f)]. All spectra exhibit an asymmetric intensity of the spectral weight at $\pm\DeltaT$, {\it i.e.}, at zero-energy. 

\section{\texorpdfstring{F\MakeLowercase{e}}~~monomers and dimers}

\begin{figure}[h]
\includegraphics[width=\columnwidth]{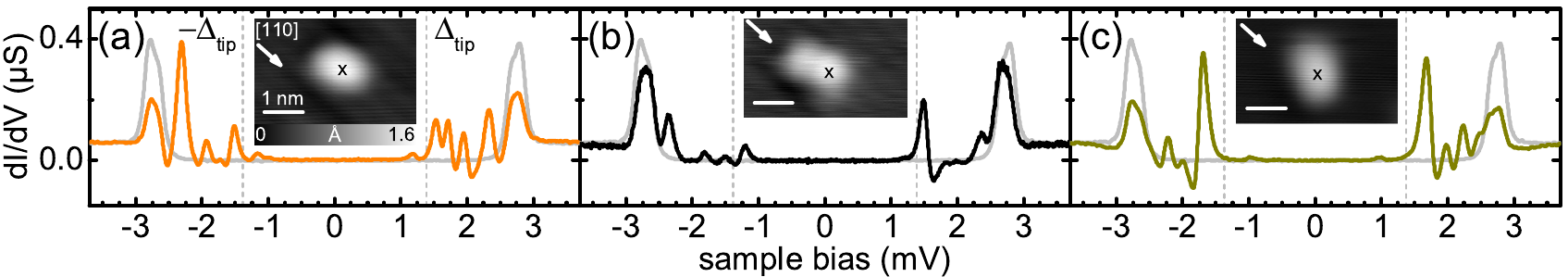}
\caption{
(a-c) \didv-spectra of various Fe dimers [reference spectrum on pristine Pb(110) in grey as guide to the eye]. The spectra show diverse subgap structures, probably due to different adsorption configurations.
(a) is the dimer shown in \MainFig{2}(b) of the main text. \DeltaT in mV: (a) 1.39, (b) 1.38, (c) 1.37.
Setpoint: $5\umV$, $250\upA$. Lock-in: $15\umuV_\mathrm{rms}$, $912\uHz$.
}
\label{Sfig:dimers}
\end{figure}

In the main text we show \didv-spectra of the two species of Fe adatoms, which are present on the (110) surface after the evaporation of  Fe onto the cold sample ($T_{sample}<10\,K$).
Both species are adsorbed in between the [110] corrugation lines of the surface, yet they have different apparent heights ($\Delta z\simeq20\upm$ at $V_{bias}=50\umV$, $I=50\upA$). In \didv spectroscopy between $\pm3.5\uV$, we did not detect any d-state resonances. We occasionally found -- besides single adatoms -- also larger protrusions  of different size and shape. 
As diffusion is hindered at the temperature of deposition, it is reasonable to assume that the vast majority of these are Fe dimers. 
We show the \didv-spectra of three distinctly different dimers in Fig.~\ref{Sfig:dimers}. All of them show a rich subgap structure, indicating a strong interaction of the Shiba states. However, the subgap structure varies strongly, and depends on the inter-atomic distance, the angle, and the adsorption site of the Fe atoms. 
\newpage
\section{\texorpdfstring{{\protect \didv}}~-spectra of the iron chain d-bands}

\begin{figure}[h]
\includegraphics[width=0.6\textwidth]{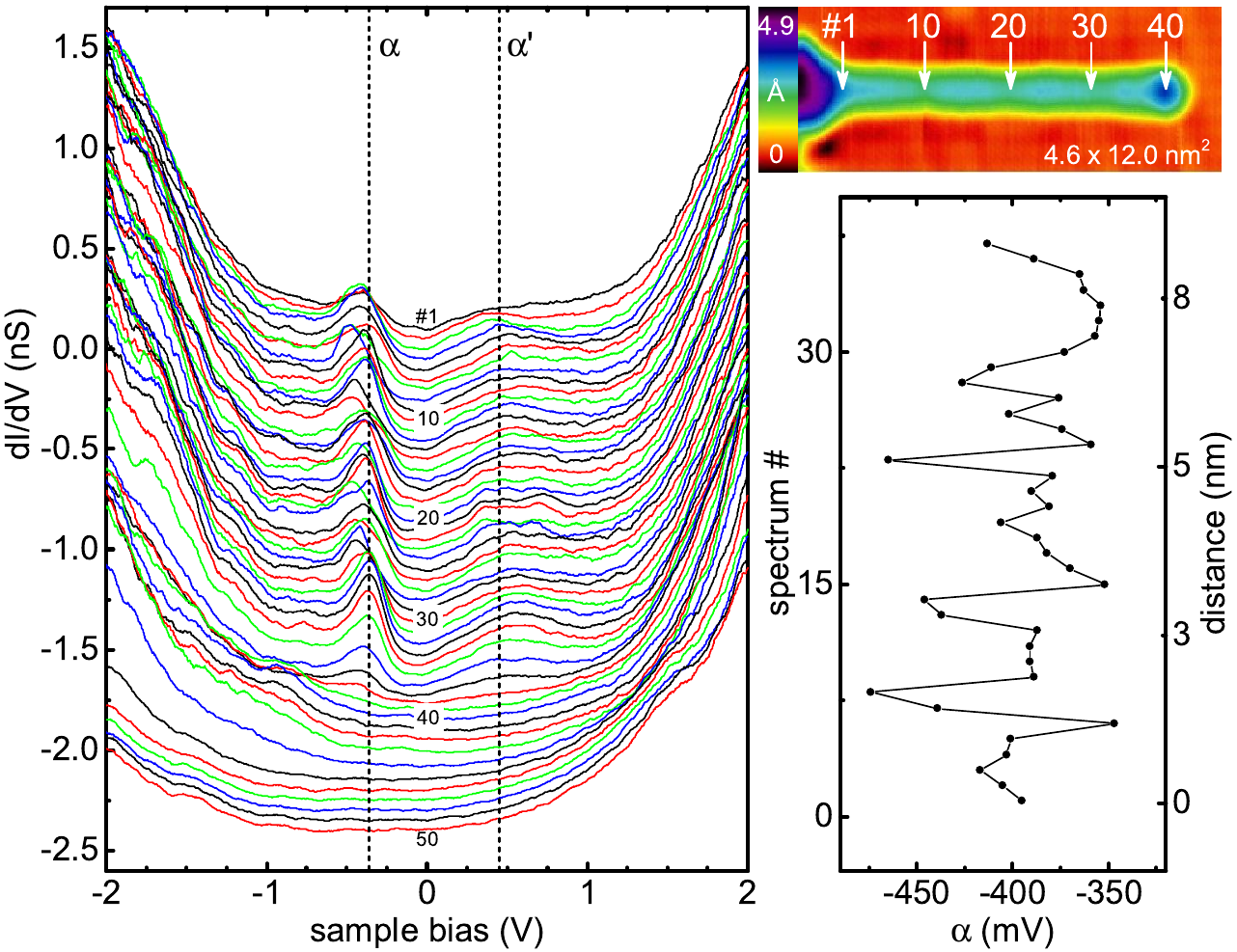}
\caption{
\didv-spectra (left) from top to bottom recorded along the chain (top right) starting from the iron cluster and going towards the chain end. The spectra are numbered from \#1 to \#50. The data is the same as plotted in \MainFig{4}(c) of the main text. Resonances $\alpha$ and $\alpha'$ are marked by dashed lines. The energetic position of $\alpha$ is plotted versus the spectrum number and the distance (bottom right). The values are determined from a fit with a Gaussian peak and a linear background. Spectra are smoothed to the adjacent average of 25 points.
Point distance $230\upm$. Offset for clarity: $-0.05\unS$/spectrum. Setpoint: $2\uV$, $850\upA$. Lock-in: $2\umV_\mathrm{rms}$, $912\uHz$.
}
\label{Sfig:dbands}
\end{figure}

Figure~\ref{Sfig:dbands} shows \didv-spectra acquired on the chain presented in \MainFig{4} of the main text. The spectra visualize the variations in intensity and energy of the resonances $\alpha$ and $\alpha'$ along the chain. At the protrusion at the chain end, these resonances have not yet developed. Instead, we observe a resonance around $-930\umV$ (spectrum \#40). It decays quickly along the chain. In spectrum (\#38) it is hardly visibly anymore, while resonances $\alpha$ and $\alpha'$ start to gain intensity.
Both increase in intensity when moving towards the center of the chain. In the center ($\sim$\#20), $\alpha'$ is resolved as double-peak structure. This correlates with the tight-binding calculations presented in Ref.~\cite{yazdani}, and may be a hint that the broad resonance $\alpha'$ actually consist out of two resonances.
The simultaneous appearance of $\alpha$ and $\alpha'$ suggests that they originate from the same band. Different positions within the chain show a shift of the resonance $\alpha$ 
by up to $150\umeV$, which is plotted in the bottom right of Fig.~\ref{Sfig:dbands}. At the Fe cluster both resonances decrease in intensity again (\#1).

\begin{figure}[t]
\includegraphics[width=0.95\columnwidth]{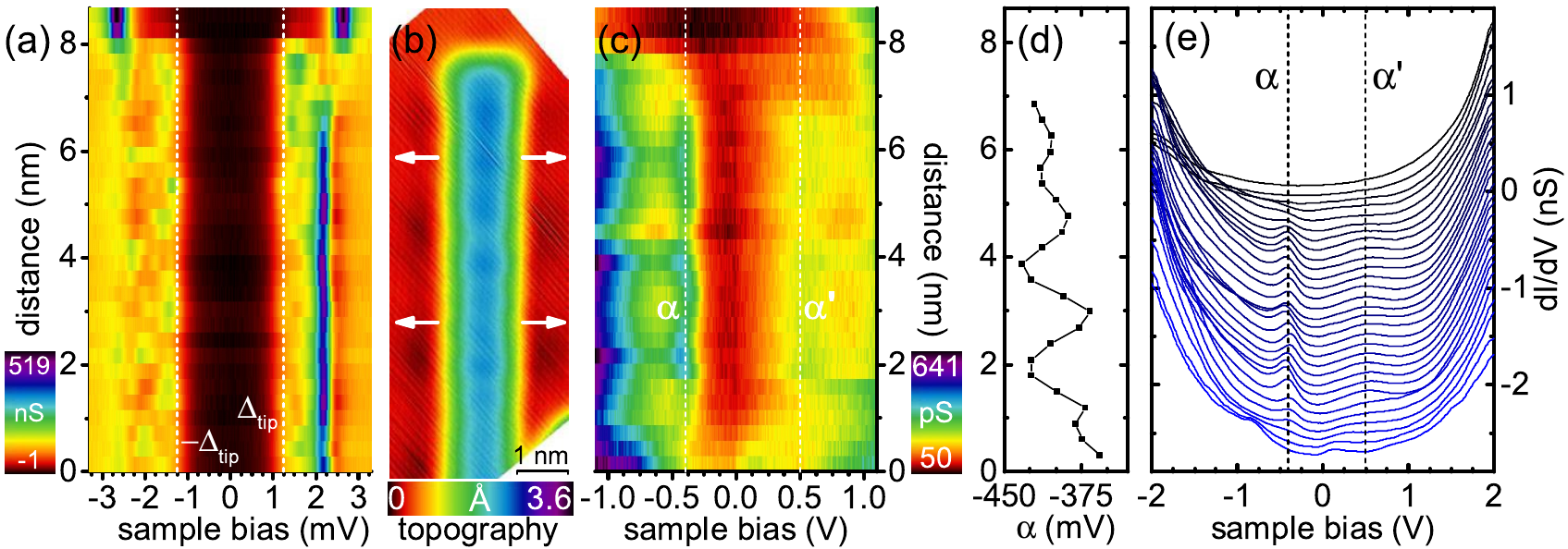}
\caption{
\didv-intensity of the subgap structure (a), and the large bias range (c) as color plot with respect to the location along the chain and the sample bias.
The topography in (b) is aligned to the position of the spectra in the color plot. The chain is the same as shown in \MainFig{1}(e) of the main text. The tip has a reduced energy resolution of $330\umuV$.
The position of $\alpha$ is plotted in (d). The values are determined from a fit with a Gaussian peak and a linear background.
(e) shows the full \didv-spectra of the data shown in (c). Data is smoothed to the adjacent average of 15 points.
Point distance $3\uAA$. Offset for clarity: $-0.1\unS$/spectrum.
Setpoint: $2\uV$, $850\upA$ for (c) and (e), and $5\umV$, $250\upA$ for (a).
Lock-in: $2\umV_\mathrm{rms}$, $912\uHz$  for (c) and (e), and $15\umuV_\mathrm{rms}$, $912\uHz$ for (a).
}
\label{Sfig:dbandsChainWithoutEnd}
\end{figure}

A similar behavior of the state $\alpha$ 
is found in a chain, which is not terminated with an Fe cluster. As an example, we show a set of spectra in Fig.~\ref{Sfig:dbandsChainWithoutEnd}. The resonances $\alpha$ and $\alpha'$ again appear simultaneously close to the end of the chain. We do not observe a resonance at larger negative bias at the end of the chain, which resembles the one at $-930\umV$ in Fig.~\ref{Sfig:dbands}. This supports the correlation of its appearance with the protrusion at the end of the chain.
Though, approaching the large Fe cluster, $\alpha$ and $\alpha'$ vanish, and resonances at $-800\umV$ and around $150\umV$ appear in the bottom of Fig.~\ref{Sfig:dbandsChainWithoutEnd}(e).
Within the chain, resonance $\alpha$ exhibits a clear oscillation in its peak position with a periodicity $\simeq 2\unm$, similar to the previously described chain. Resonance $\alpha'$ also shows a variation in its peak position, but not with a clearly identifiable periodicity. We also present the corresponding subgap structure in Fig.~\ref{Sfig:dbandsChainWithoutEnd}(a). The most prominent Shiba state at $2.17\umV$ shows an intensity oscillation, in accordance with the variation of $\alpha$. Due to the limited energy resolution in these spectra, we cannot unambiguously conclude on an energy variation.

\newpage
\section{Determination of the tip gap}

We use superconducting tips in order to improve the energy resolution beyond the Fermi-Dirac limit at $1.1\uK$ and to detect asymmetries in the electron and hole components of the subgap states.
\didv-spectra show thus the measured spectral intensity of the sample convolved with the BCS-like density of states of the tip. A consequence of a superconducting tip with gap \DeltaT is the shift of a sample resonance with an energy $\pm\varepsilon$ to a bias value of $\pm\left(\DeltaT\pm\varepsilon\right)/e$. 
The exact determination of ``real'' energies of the sample resonances thus relies on the correct determination of the superconducting gap of the tip \DeltaT.

Pb is a two-band superconductor with two gap parameters ($\Delta_1\simeq1.42\umeV$ and $\Delta_2\simeq1.27\umeV$). They originate from two separated Fermi surfaces, and give rise to the double-peak structure in the \didv-spectra~\cite{ruby14}. The tip is prepared by controlled indentation into the clean Pb surface with a high voltage applied to the tip. This creates an amorphous superconducting Pb layer on the tip and yields a single gap parameter \DeltaT, which is averaged over all directions. Depending on the layer thickness and quality,  \DeltaT can be similar or smaller than the bulk gap values.

We cannot determine the parameters $\Delta_1$, $\Delta_2$, and \DeltaT independently from the BCS resonances in the spectra of pristine Pb(110) alone.
An independent determination of the full set of parameters ($\Delta_1$, $\Delta_2$, and \DeltaT) is only possible using spectra with a pronounced low-energy Shiba state, which gives rise to well-resolved thermal resonances. Fe dimers show such low-energy Shiba resonances (Fig.~\ref{Sfig:dimers}). 
The Shiba resonance and its thermal counterpart occur symmetric to \DeltaT at $\pm\left(\DeltaT+\varepsilon\right)$ and $\pm\left(\DeltaT-\varepsilon\right)$, respectively. 
This allows us to determine \DeltaT unambiguously. Spectra of the pristine surface acquired with the same tip show clear BCS resonances at $\DeltaT+\Delta_{1,2}$. Because $\Delta_1$ and $\Delta_2$ are bulk properties of the substrate, their energy can then serve to determine \DeltaT  for every tip. 
This procedure enables a reliable determination of the energies of subgap resonances in each Fe chain.

%
%
\newpage

\section{Theoretical Model}

We use the model introduced in Ref.~\cite{peng2015}, where the Fe adatoms are modeled as a chain of Anderson impurities hybridizing with the substrate BCS superconductor. The only difference is that we account for the corrugation of the chain by assuming a spatially varying on-site energy of the Anderson levels. We choose parameters such that only one spin band, say spin down, crosses the Fermi level, whereas the spin-up band is far below the Fermi level. As described in Ref.~\cite{peng2015} we can evaluate the Green function within mean-field theory, from which we numerically obtain the local density of states (LDOS) at subgap energies. In our numerical calculations, the average on-site energy of the spin-down levels is $\overline{E}_{d,\downarrow}=200\Delta_{\mathrm{sample}}$, with roughly $10\%$ spatial modulation (see Fig.~5(b) of the main text). The onsite energy of the spin-up states is $E_{d,\uparrow}=E_{d,\downarrow}-40000\Delta_{\mathrm{sample}}$. We choose the bandwidth of the spin-down band as $1000\Delta_{\mathrm{sample}}$, the hybridization strength between the adatoms and the substrate superconductor $\Gamma=256\Delta_{\mathrm{sample}}$, and the ratio $\Delta_{\mathrm{sample}}/\Delta_{\mathrm{tip}}=0.958$. 
Finally, we set the Fermi wavevector $k_F a=4.3\pi$ and the Rashba spin-orbit wavevector $k_h a=0.26\pi$ with $a$ the lattice constant.

\begin{figure}[bt]
    \includegraphics[width=0.9\textwidth]{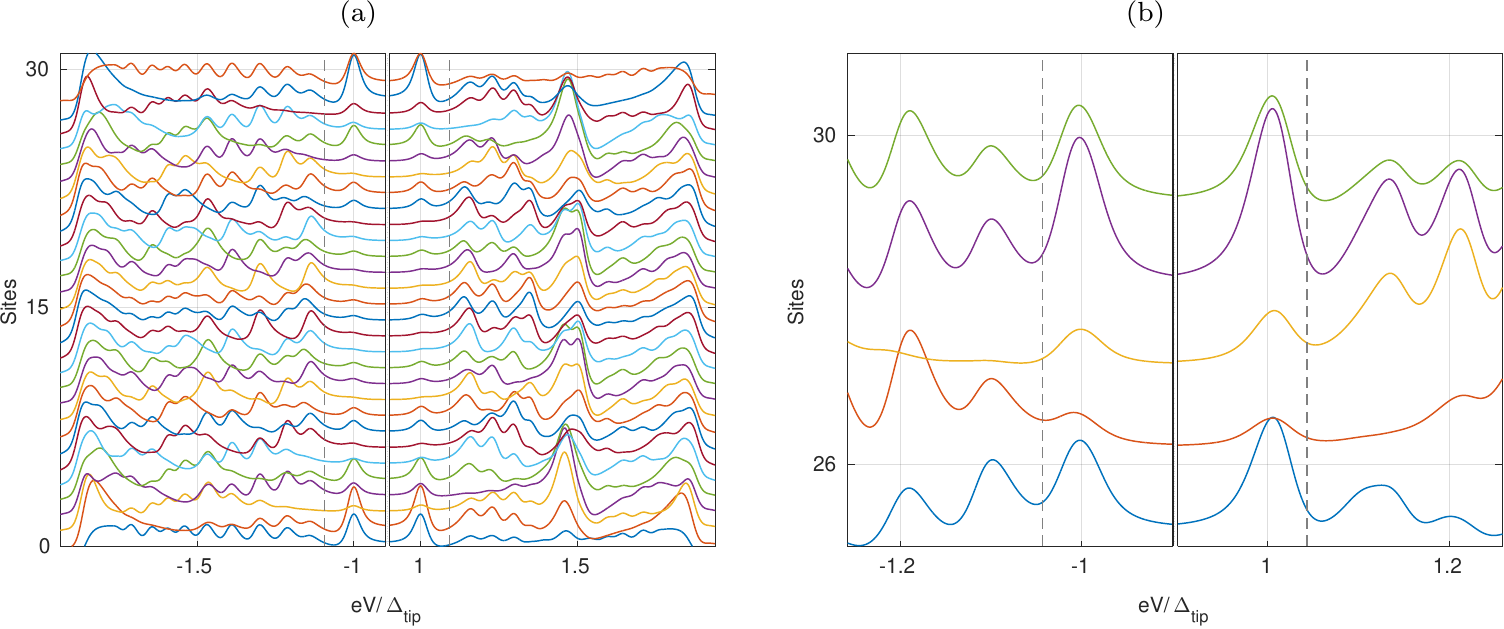}
    \caption{Numerical results for the differential conductance measured with a superconducting tip at subgap energies along a chain of $30$ sites. (a) Sample with a larger
        $p$-wave induced gap with $k_F a=4.3\pi$, same as the color scale plot in Fig.~5(a) of the main text. (b) Sample with a smaller $p$-wave induced gap with $k_F a=8.3\pi$, 
        only the last 5 traces around $\Delta_{\mathrm{tip}}$ are shown. 
        The dashed lines indicate $eV=\pm(\Delta_{\mathrm{tip}}+\Delta_{p})$, with $\Delta_{p}$ the induced $p$-wave gap for an infinite chain. In (a) $\Delta_{p}=0.096\Delta_{\mathrm{tip}}$ and in (b) $\Delta_{p}=0.045\Delta_{\mathrm{tip}}$. The remaining model parameters and the process of evaluating the conductance are the same.}
	\label{Sfig:waterfall}
\end{figure}

\begin{figure}[bt]
    \includegraphics[width=0.9\textwidth]{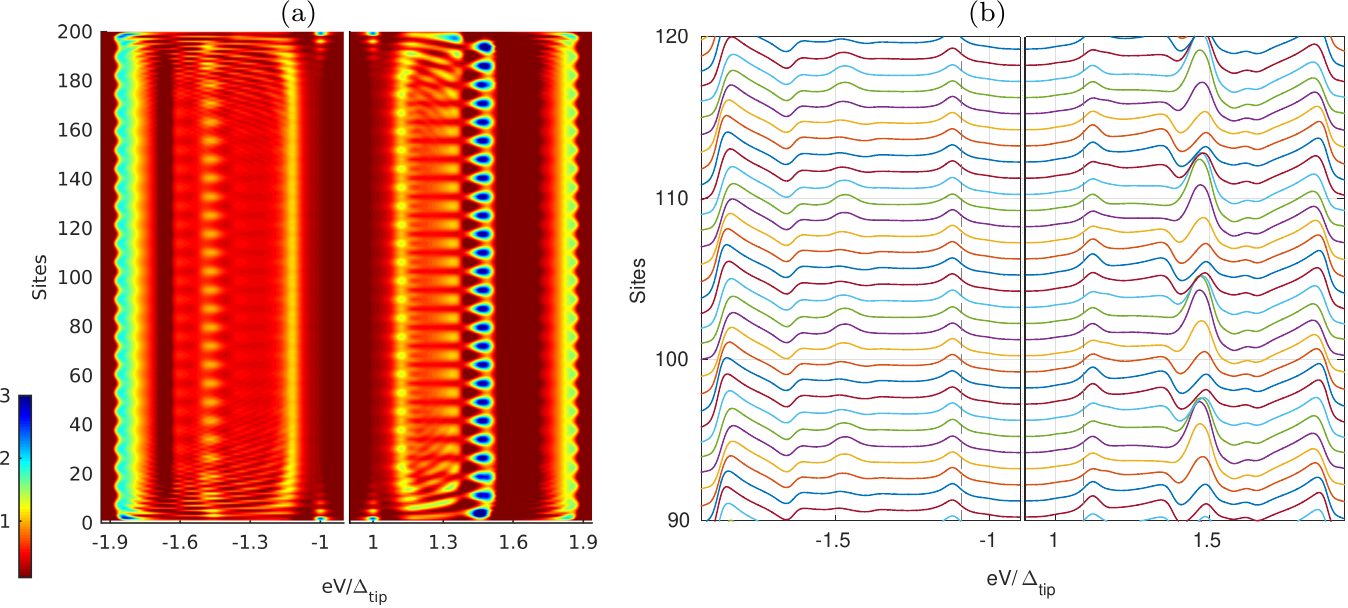}
    \caption{Numerical results for the differential conductance measured with a superconducting tip at subgap energies along a longer chain of $200$ sites with the same parameters as in Fig.~\ref{Sfig:waterfall}(a), in particular the same modulation in the on-site energy of impurities. (a) Color scale plot of the conductance. (b) Waterfall plot for $30$ sites in the middle of the chain. The dashed lines indicate the bias $\pm(\Delta_{\mathrm{tip}}+\Delta_{p})$ corresponding to the $p$-wave gap of an infinite chain.}
        \label{Sfig:len_200}
\end{figure}

We start from the expression for the tunneling current (see supplement of Ref.~\cite{yang2}), 
\begin{equation}
    I=\frac{et^{2}}{2h}\int d\omega\,\Tr\left\{ G_{R}^{>,ee}(\omega)g_{L}^{<}(\omega_{-})-G_{R}^{<,ee}(\omega)g_{L}^{>}(\omega_{-})+g_{L}^{>}(\omega_{+})G_{R}^{<,hh}(\omega)-g_{L}^{<}(\omega_{+})G_{R}^{>,hh}(\omega)\right\},
\end{equation}
where $R$ and $L$ label the sample and the superconducting tip and all Green's functions are located at the tunneling position. The electron and hole blocks are denoted by $ee$ and $hh$. In the experiment, the tunneling current is dominated by single-particle tunneling events (cf.\ Ref.~\cite{ruby14}) and we can approximate
\begin{gather}
    -iG_{R}^{<}(\omega)=A(\omega)f(\omega),\\
    iG_{R}^{>}(\omega)=A(\omega)(1-f(\omega)), 
\end{gather}
where $f(\omega)$ is the quasi-equilibrium distribution of the steady state and $A(\omega)=-2{\rm Im}G_{R}^{r}(\omega)$ the LDOS of the sample. The distribution $f(\omega)$ satisfies the condition $f(-\omega)=1-f(\omega)$, as in the case of the Fermi distribution function. For our numerical calculations, we assume that the system remains close to thermal equilibrium, $f(\omega)=n_F(\omega)$. Using the relations for the tip Green function $g_{L}^{<}(\omega)=2\pi i\rho(\omega)n_{F}(\omega)$, $g_{L}^{>}(\omega)=-2\pi i\rho(\omega)n_{F}(\omega)$, we can write the current as
\begin{align}
    I&=\frac{\pi et^{2}}{h}\int d\omega\,
        \rho(\omega_{-})    \Tr A_{e}(\omega)\left[n_{F}(\omega_{-})-f(\omega)\right]-\rho(\omega_{+})\Tr A_{h}(\omega)\left[n_{F}(\omega_{+})-f(\omega)\right] \\
                &=\frac{2\pi et^{2}}{h}\int d\omega\, \rho(\omega_{-})\Tr A_{e}(\omega)\left[n_{F}(\omega_{-})-f(\omega)\right],
\end{align}
where $A_{e,h}$ denotes the $2\times 2$ electron (hole) block of $A$. 
In the last line, we have used $\Tr A_{e}(\omega)=\Tr A_{h}(-\omega)$ due to particle-hole symmetry, and  
the property $f(-\omega)=1-f(\omega)$.
The conductance is then obtained by taking the derivative with respect to the voltage, 
\begin{equation}
G(eV)=-\frac{2\pi e^{2}t^{2}}{h}\int d\omega\Tr\{A_{e}(\omega)\}\left\{ \rho'(\omega_{-})\left[n_{F}(\omega_{-})-f(\omega)\right]+\rho(\omega_{-})n_{F}'(\omega_{-})\right\}, 
\label{eq:GeV_p}
\end{equation} 
where $\rho'(\omega)=d\rho(\omega)/d\omega$, and $n_{F}'(\omega)=dn_{F}(\omega)/d\omega$. 
Inverting the sign of the bias voltage, we find
\begin{equation}
    G(-eV)=-\frac{2\pi e^{2}t^{2}}{h}\int d\omega\Tr\{A_{e}(-\omega)\}\left\{ \rho'(\omega_{-})\left[n_{F}(\omega_{-})-f(\omega)\right]+\rho(\omega_{-})n_{F}'(\omega_{-})\right\} .
    \label{eq:GeV_m}
\end{equation}
Since in general $\Tr A_{e}(\omega)\neq\Tr A_{e}(-\omega)=\Tr A_{h}(\omega)$, the conductance is not necessarily symmetric under reversal of bias voltage. However, an isolated
zero-energy resonance yields symmetric peaks at $eV=\pm \Delta_{\mathrm{tip}}$ \cite{yang2}. In Fig.~\ref{Sfig:waterfall}(a) we show a waterfall plot of the differential conductance along the chain for the same parameters as the color plot in Fig.~5(a) of the main text. The dashed line indicates the $p$-wave gap for an infinite chain. The lowest-energy resonance in the finite chain is somewhat higher and varies along the chain due to finite size effects. In Fig.~\ref{Sfig:waterfall}(b) we show that a smaller $p$-wave gap leads to an asymmetry at $eV=\pm\Delta_{\mathrm{tip}}$. In Fig.~\ref{Sfig:len_200} we show the conductance plot for a longer chain, where the lowest-energy resonance is now more homogenous along the chains and closer to the induced gap. The small offset from $eV=\pm(\Delta_{\mathrm{tip}}+\Delta_{p})$ originates from the energy-dependent density of states in the superconducting tip and is not due to finite-size effects.

\end{document}